\documentclass[12pt,3p]{article}
\usepackage{jheppub}
\usepackage{amsmath,amssymb,latexsym}
\usepackage{url}
\usepackage{float}
\usepackage{graphics}
\usepackage{graphicx}
\usepackage{epsfig}
\usepackage{psfrag}
\usepackage{color}
\usepackage{slashed}

\usepackage{amsfonts}

\begin{document}

\title{KK Higgs produced in association with
	a top quark pair in the bulk RS Model}

\author{N. Manglani$^{a,b}$,} 
\author{A. Misra$^b$,}
\author{K.Sridhar$^c$}
\affiliation[a]{Shah and Anchor Kutchhi Engineering College, Mumbai 400088, India}
\affiliation[b]{Department of Physics, University of Mumbai, Kalina, Mumbai 400098, India}
\affiliation[c]{Department of Theoretical Physics, Tata Institute of Fundamental Research, Homi Bhabha Road, Colaba, Mumbai 400 005, India}

\emailAdd{namrata.manglani@sakec.ac.in}
\emailAdd{misra@physics.mu.ac.in}
\emailAdd{sridhar@theory.tifr.res.in}

\abstract{We present a search strategy for the first Kaluza-Klein (KK) mode of the Higgs boson in the framework of the 
	Randall-Sundrum (RS) model with a deformed metric. We study the production of this massive excitation in 
	association with a $ t \bar t$ pair at the Large Hadron Collider (LHC). The KK Higgs primarily decays into
	a boosted $t \bar t$ final state and we then end up with an interesting four-top final state of which 
	two are boosted. The boosted products in the final state improve the sensitivity for the search of the KK 
	Higgs in this channel whose production cross-section is otherwise rather small. Our results suggest that 
	masses of the KK Higgs resonance upto about 1.2 TeV may be explorable at the highest planned luminosities 
	of the LHC. Beyond this mass, the KK Higgs cross-section is too tiny for it to be explored at the LHC and 
	may be possible only at a future higher energy collider.
	}

\maketitle

\section{Introduction}
 
 The Randall-Sundrum (RS) model of warped extra dimensions addresses issues of gauge and mass hierarchy in
 a simple manner and may yet provide some new observable consequences at the Large Hadron Collider (LHC). The
 model, as originally conceived, provides a neat separation of gravity and Standard Model (SM) physics by
 localising these on two branes (UV and IR) located at the end-points of a segment of a five-dimensional bulk, 
 respectively. The localisation of all SM particles on the IR brane has disastrous consequences for the model:
 the IR localised fields turn out to be composite and the model, consequently, fails miserably when
 confronted with electroweak precision measurements. One has to differentially localise
 the various SM fields in the bulk with only the Higgs field localised on or near the IR brane and the models
 thus constructed are what are known as the Bulk RS Models \cite{Raychaudhuri:2016,Gherghetta:2010cj,Davoudiasl:1999tf,Gherghetta:2000qt,Pomarol:1999ad,Grossman:1999ra,Chang:1999nh}.
  But even this is not enough to ensure agreement 
 with electroweak measurements and one has to invoke a bulk custodial symmetry  or to deform the AdS 
 metric near the IR brane leading to two different Bulk Models: the custodial model \cite{Agashe:2003zs} and the deformed model \cite{Cabrer:2011fb}.
 
 Even after making these modifications both the models demand that, the Kaluza-Klein (KK) modes of the
 bulk-localised SM particles which are constrained by electroweak data should be rather heavy. In the custodial model,
 these turn out to be of the order of 2.5-3 TeV, at least, for the first KK mode but, generically, the
 deformed model allows for lighter KK resonances. It is possible to accommodate a KK Higgs in the deformed 
 model in the 1 TeV range and it is the KK Higgs in this mass range that will interest us in the rest
 of this paper. In other words, we are restricting ourselves to the deformed model and discussing KK Higgs
 production in the context of this model. In an earlier paper, we have discussed, in the context of the
 deformed model, the production of a KK Higgs resonance in gluon-gluon fusion through a triangle of top quarks \cite{Mahmoudi:2017txo}, 
 In this paper, we study KK Higgs production in association with a $t \bar t$ pair. This, as we know from 
 even the SM case, is a difficult mode to study because of the tiny cross-sections. Our attempt is to see 
 how much of the KK Higgs mass range can be covered in the future runs of the LHC. At the outset, our 
 intuition that this is not a hopeless task is built on the observation that the KK Higgs decays also into 
 a boosted top-pair and the four-top final state with two tops boosted is shielded from
 very large SM backgrounds. Having said that, this is by no means an easy task and we devote the rest of the
 paper to describing what we can do to enhance the signal efficiency and have a reasonable number of events
 in this channel.

  The paper is organised as follows: In section 2  we have briefly discussed direct constraints on the four top final state obtained from colliders. Section 3 contains details of the signal and background simulation and the search strategies proposed to enhance the signal and achieve relevant background rejection. Finally in section 3 we present results and our conclusion is given in section 4.

 \section{Constraints on a four top final state search}
 
 The four-top final state studies at the colliders are rare due to the complexity of the final state and its very low cross sections. Even at the top factory like the LHC, the SM cross section for a four top final state is negligible (11 fb) in comparison to the huge SM two top background.  Searches for such a final state have been done in Refs. \cite{Khachatryan:2014sca,Alvarez:2016nrz,Kim:2016plm,Aaboud:2018xuw}. We refer to most recent work by Aaboud et. al. which provide the most stringent bounds. They put a limit on the production cross section of $16 ^{+12}_{-9} fb $ on a massive resonance produced in association with top-quark pair resulting into a four-top final state. 
 The cross-section for the process that we are considering is much smaller than the aforementioned limit and the KK-Higgs production does not show up at current luminosity. Even at higher luminosity the search has to be carefully conducted and this is what we now discuss.
 
 \section{Signal and Background simulations }
 A top-quark decays to a W boson and a b-quark, where a W boson can decay either to a lepton and neutrino or two quarks . Thus for any four-top final state we will have a 12 parton final state, which should have a minimum of 4 b-quarks. Based on the most possible combinations, the final state is classified into four categories \cite{Alvarez:2016nrz} namely, hadronic (All top-quarks decay hadronically), semileptonic (one of the four top-quarks decays leptonically), similiarly the dilepton and trilepton final states are where two and three top-quarks decay leptonically respectively.
 
 We consider a special four-top final state:
 \begin{equation}
 p p\rightarrow H_{1} t\bar t \rightarrow t\bar t t\bar t.
 \end{equation}

In the above process the four-top final state is special in a sense that two of the top-quarks that are decay products of $H_{1}$ are boosted. Such a differential distribution of boost among the four top-quarks in the final state makes our search interesting to explore. 

For this special four-top final state we find that, the case where two boosted top-quarks decay hadronically, is the only one where, efficient mass reconstruction of $H_{1}$ is possible, provided, other irreducible backgrounds are efficiently handled. We in the present work have performed a thorough inspection of this special four-top final state signal, where, two boosted top-quarks decay hadronically, with a thrust on selective reduction of multijet QCD backgrounds. We select events which do not have a lepton with $p_{T} > $ 25 GeV in order to narrow down on events where boosted top-quarks decay hadronically. So we find that the special four-top final state that we are left with is either hadronic (all four top-quarks decaying hadronically) or special semileptonic, where, boosted top-quarks decay hadronically and the lepton that comes from one of two spectator top-quarks has a $p_{T}$ less than 25 GeV \footnote{Lepton isolation is not used here as we found it to reduce signal efficiency in the crowded four-top final state}.

We closely follow Ref. {\cite{Kim:2016plm}} for the choice of backgrounds. The Standard Model four-top and the top-quark pair with two additional heavy quark jets form  two irreducible backgrounds for our signal. Their cross sections  at 14 TeV centre of mass energy are given in the Table \ref{bg_table} .  Other Multijet backgrounds that have not been listed from Ref. \cite{Kim:2016plm} in the following table are QCD multijet background, $t\bar{t}$+jets \footnote{Jets excluding b-quark jets}, $t\bar{b}$+jets and $b\bar{b}$+jets. We are able to tab them by our choice of cuts, what still remain substantial post cuts are the irreducible backgrounds given below:

 \begin{table}[h!]
 	
 	\begin{center}
 		\begin{tabular}{|c|c|c|l|} \hline
 			Sr. No.& Backgrounds & Cross sections \\ \hline
 		
 			1& $t\bar{t}t\bar{t}$ &11.81 fb\\
 			2& $t\bar{t}bb$&16.5 pb \\
 	
 			\hline
 		\end{tabular}
 	\end{center}
 	\caption{ The simulated cross sections for the irreducible SM backgrounds for the hadronic decay channel in the four top final state .}\label{bg_table}
 \end{table}
 
 In Table \ref{bg_table}, $t\bar{t}t\bar{t}$  is the SM four top final state and $t\bar{t}bb$ is another mutijet background. The parton-level amplitudes for both the signal and background were generated using {\tt{MADGRAPH}} \cite{Alwall:2014hca} at 14 TeV centre of mass energy using 
 parton distribution function {\tt{NNLO1}} \cite{Ball:2012cx} and subsequently showering is done in {\tt{PYTHIA}} \cite{Sjostrand:2014zea} while the bulk higgs model files have been generated using {\tt{FEYNRULES}} \cite{Alloul:2013bka}.

 Once the simulation of the signal and background is completed using {\tt{MADGRAPH}} and  {\tt{PYTHIA}}, then all final state particles in the signal and background are clustered into jets employing the $anti-k_T$ \cite{Cacciari:2008gp} clustering algorithm in {\tt{FASTJET}} \cite{Cacciari:2006sm,Cacciari:2011ma} with jet radius parameter set to $R=0.4$. We accept only those jets which satisfy $|\eta|<2.7$ and $p_T>40$ GeV.\\
 
 As two top quarks are resulting from the decay of massive $H_{1}$ they are boosted in comparison to other two top quarks. This as mentioned earlier is a unique feature of our signal. Along with this other features that are special for any four top final state in general are \cite{Khachatryan:2014sca}:
 
 \begin{itemize}
 	\item $N_{jets}$ : The jet multiplicity or the total number of jets. This variable generally has a large value for a four-top final state, as jets are formed from a mimimum of 12 partons. We expect $N_{jets}$ to be atleast greator than 9 for any four- top final state.
 	\item $N_{btags}$ : The b-jet multiplicity the total number of b-tagged jets. A four top final state should by virtue have four b-quarks.  Due to the limitation of b-tagging efficiency we can tag a minimum of three b-quarks per event and still have sufficient number of signal events passing this cut. So we expect $N_{btags}$ greator than equal to 3.
 	\item $H_{T}$ : The total scalar sum of $p_{T}$ for all the jets. As the $H_{1}$ decays to boosted top-quarks the resulting decay products of such top-quarks are also boosted. Hence, $p_{T}$ of jets is high in our case, thus this total sum $H_{T}$  is very high. The maximum cut on $H_{T}$ can be  around 1200 GeV for $M_{h_{1}}$= 1TeV, for higher mass of $H_{1}$ this value will be still higher.
 	
 \end{itemize}
 
\begin{figure*}
 	\begin{center}
 		\begin{tabular}{cc}	
 			\includegraphics[width=7.5 cm, height= 6cm]{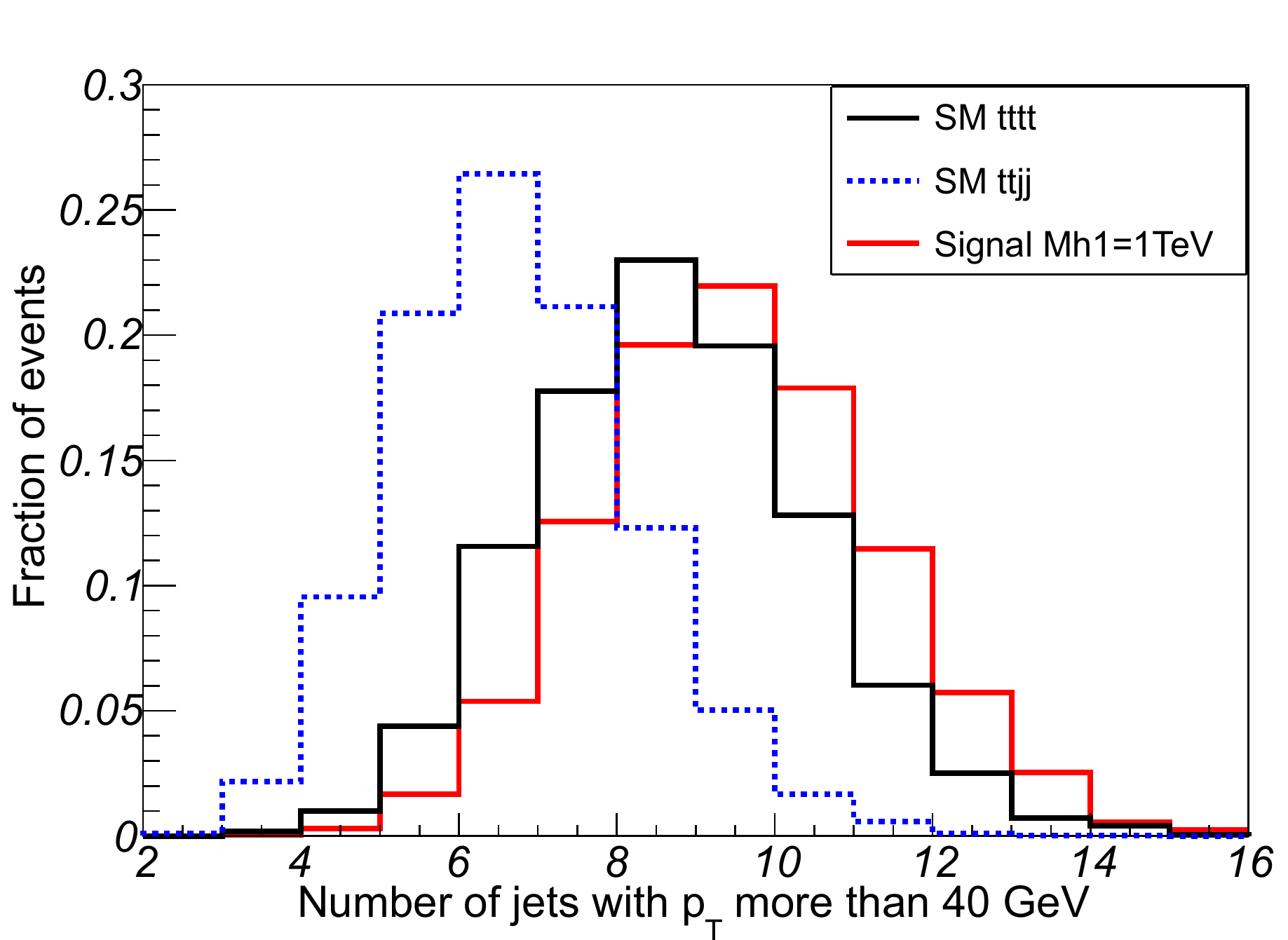}&	
 			\includegraphics[width=7.5 cm, height= 6cm]{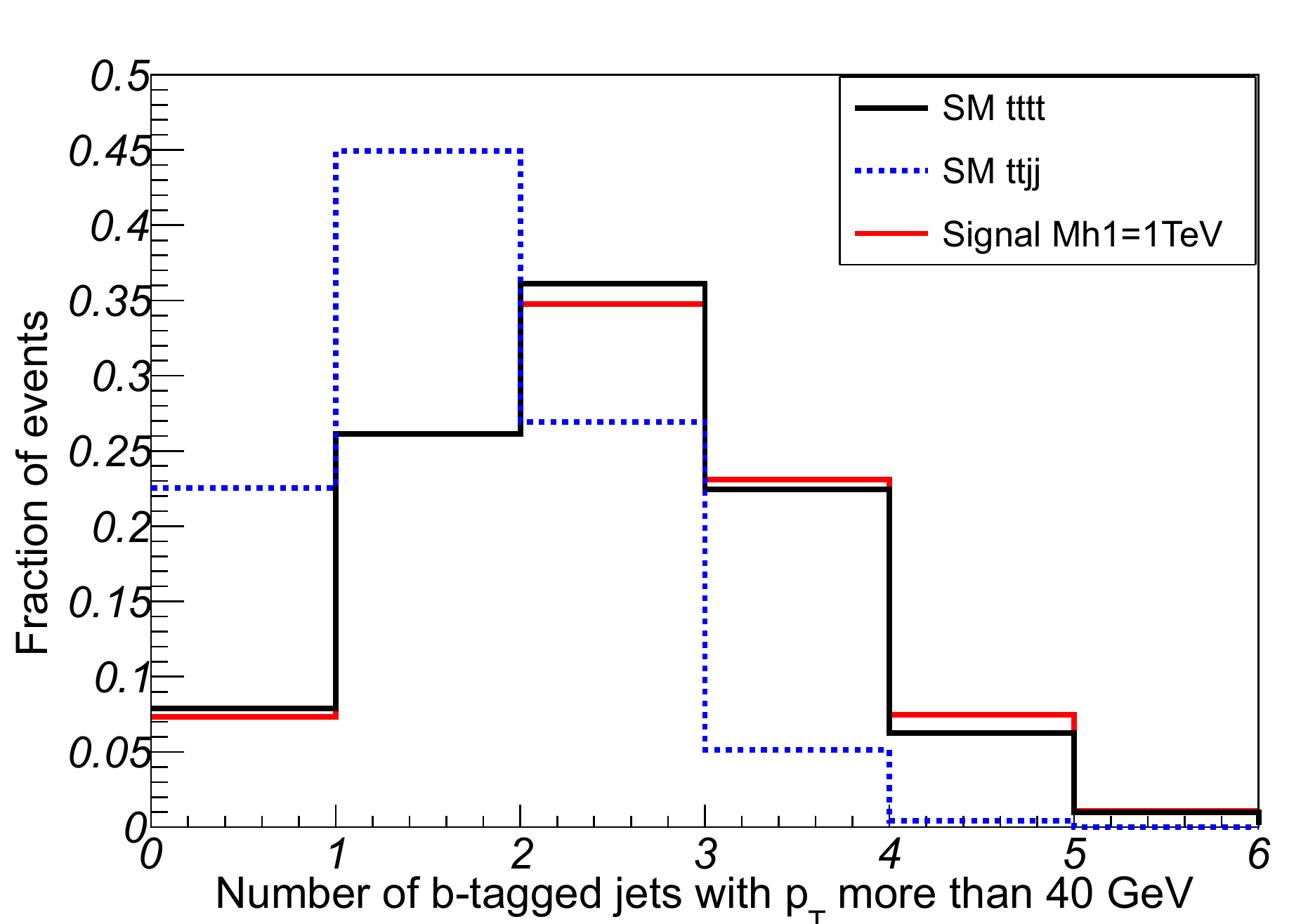}\\
 		\end{tabular}
 		\caption{ Number of jets to the left and the number of b-tagged jets as a discriminating variable to the right}
 		\label{njets}       
 	\end{center}
 \end{figure*}

The above variables are the discriminating variables between our signal and the backgrounds. In Fig. \ref{njets} we show the plot for total number of jets after clustering and passing the set criteria along with number of jets that are b-tagged for our signal and all irreducible backgrounds. Our finding that the signal can be efficiently separated if the number of jets required are greater than equal to nine and b-tagged jets are greater than or equal to 3 can be clearly seen in the plots.
 
\begin{figure*}
 	\begin{center}
 		\begin{tabular}{cc}	
 			\includegraphics[width=7.5 cm, height= 6cm]{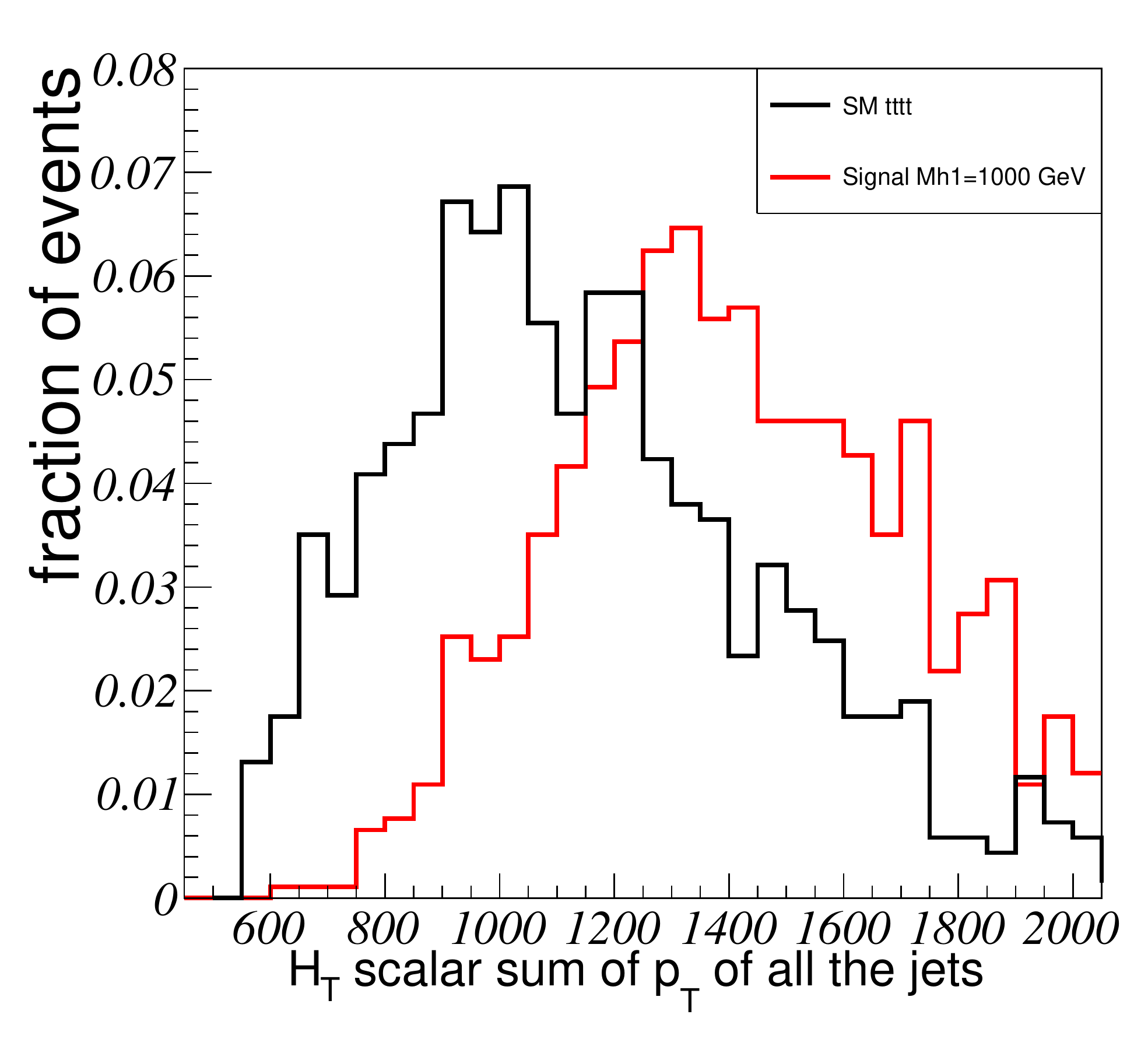}&	
 			\includegraphics[width=7.5 cm, height= 6cm]{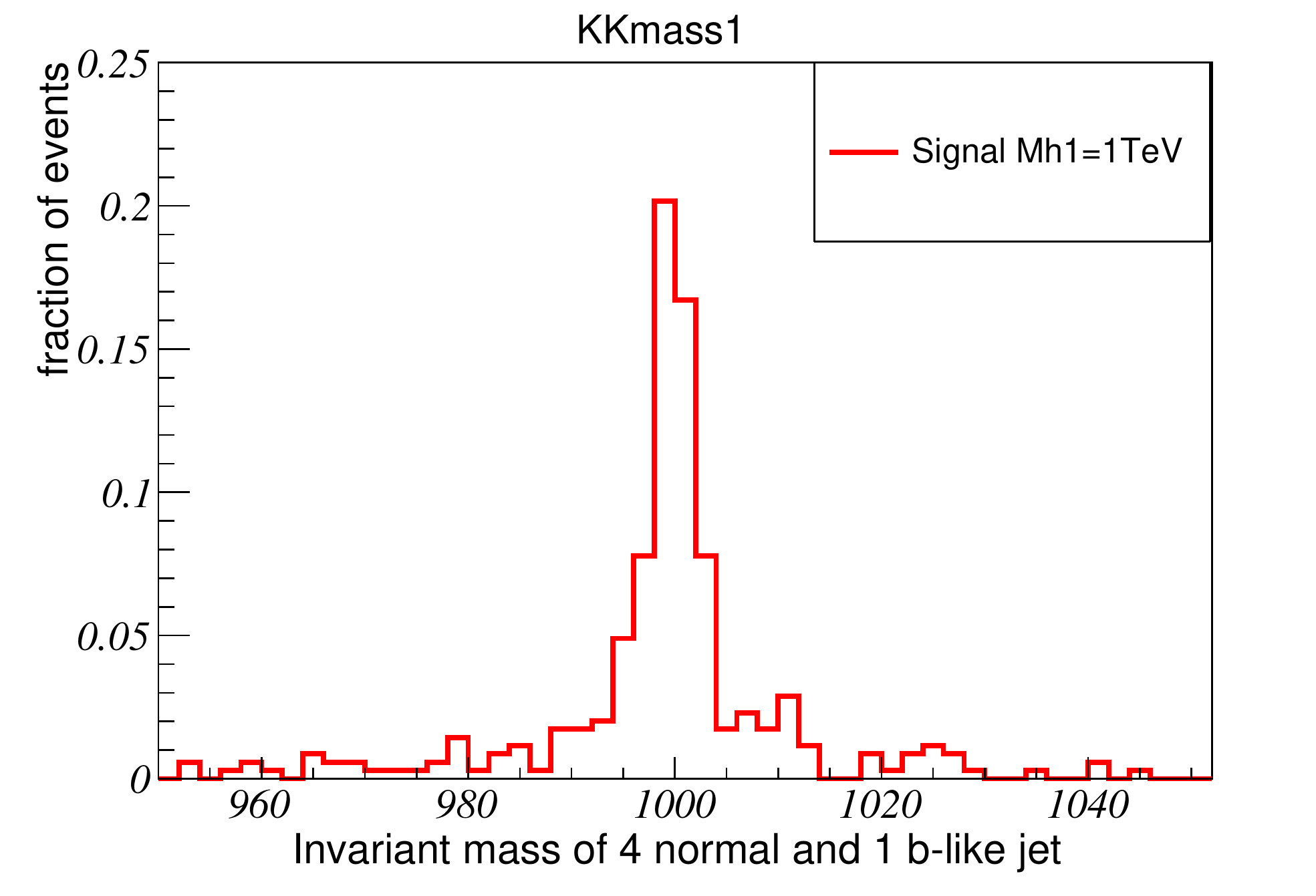}\\
 		\end{tabular}
 		\caption{ Scalar sum of $p_{T}$ of jets to the left and the $H_{1}$ mass reconstruction to the right}
 		\label{fig_htj}       
 	\end{center}
 \end{figure*}

 Futher, as mentioned earlier, in a state of high boosted jet multiplicity, another variable that comes handy for segregation of the signal is $H_{T}$. $H_{T}\geq$ 1250 GeV is the best choice for the zero lepton case that we have considered. The plot for this variable is given in Fig. \ref{fig_htj}.
 
 When $H_{1}$ decays to a pair of boosted top quarks, each top quark further decays to three partons, thus we ideally would expect six partons to result into at-least a minimum of six high $p_{T}$ jets after clustering. These jets could then be used to reconstruct the $H_{1}$ mass. Contrary to our initial expectation we find that the leading jet is highly boosted and seems to accommodate two partons hence we are able to reconstruct the $H_{1}$ mass with a total of five jets where four are normal jets and one is a b-tagged jet as shown in Fig. \ref{fig_htj}. Finally, we put a mass window cut around the $M_{h_{1}}$\\
 
 In Table \ref{l0} we show the cut flow table for the signal with KK-Higgs mass $M_{h_{1}}=1000$ GeV and the irreducible backgrounds \footnote{$t\bar t$+ jets background is replaced by $t\bar t b b$, this is because when we ask for 3 b-tagged jets, any of the events with $t\bar t + light flavour jets$ are not able pass the cuts in the senario where mistagging is not considered}.

 \begin{table}[h!]
 	
 	\begin{center}
 		\begin{tabular}{|c|c|c|c|} \hline
 			Cuts						& Signal & $ t\bar t t\bar t$  &  $t\bar t$+$b\bar b$ \\ \hline
 			
 			$N_{lepton}$=0&1.08&	8.66&	14562.70\\
 			$N_{jets}\geq$9	&0.66&	3.65&	140.35\\
 			$N_{btags}\geq$3&0.16&	0.85&	2.32\\
 			$H_{T}\geq$1250 GeV	&0.12&	0.46&	1.16\\
 			900$\leq M_{h_{1}} \leq$ 1020 GeV&0.09&	0.25&	0.50\\
 			\hline
 		\end{tabular}
 	\end{center}
 	\caption{ Cut flow table for hadronic decay of all top quarks for the $M_{h_{1}}=1 $ TeV.}\label{l0}
 \end{table}

\section{Results}
 
 Using the sequence of cuts described in the previous section for separation of the signal and the background, we obtain results for required luminosity to get a statistically significant result. These are given in Table \ref{table2}.
 
\begin{table}
	\begin{center}
		\begin{tabular}{|c|c|c|} \hline
			$Mh_{1}$& Luminosity in fb$^{-1}$ &Luminosity in fb$^{-1}$\\
			(GeV)&    	for 5$\sigma$ result &	for 3$\sigma $ result\\   \hline 
		
			900	&1046&	377\\
			1000&	2276& 819\\
			1100&	3119&	1123\\
			1200&	4008&	1443\\
		
			\hline
		\end{tabular}
	\end{center}
	\caption{Integrated luminosity in fb$^{-1}$ for 5$\sigma$ and 3$\sigma$ sensitivities.}\label{table2}
\end{table}

We have focused only on the hadronic channel and we find that in this channel the search luminosity requirement is high. We find for $M_{h_{1}}$	= 900 GeV the 5$\sigma$ luminosity is 1046 $fb^{-1}$ and $M_{h_{1}}$	= 1000 GeV the 5 $\sigma$ luminosity is 2276 $fb^{-1}$ using our proposed search strategy. Our results suggest that future higher luminosity runs may be sensitive, at least at the 3 $\sigma$ level, to the production of $H_{1}$.

In the present paper we have not performed a full detector level simulation. However the luminosity requirements may increase a little once we take into account the mistag rate in the QCD backgrounds.  

 \section{Conclusion}

In this paper, we have presented an effective search strategy for the $H_{1}$, the first KK mode of Higgs, in the context of the deformed Randall-Sundrum model. We have considered the production of $H_{1}$ in association with a $t \bar{t}$ and have studied effective ways of unravelling this special four-top final state with two boosted top-quarks. We have focused only on the hadronic channel and we find that in this channel the search luminosity requirement is high, as shown in Table \ref{table2}. We find that the luminosity requirement increases for higher masses, hence we limit our search to $H_{1}$ mass of only 1200 GeV. For higher masses cross sections are extremely tiny and have no room for exploration at the existing colliders. This is also the reason that we have not been able to use this channel to study $H_{1}$ production in the custodial model. We have to wait for future colliders that may promise higher luminosity in order to explore higher masses. 

\section*{Acknowledgements}

 N.M. would like to thank the Department of Theoretical Physics, TIFR for computational resources.

\bibliographystyle{JHEP}
\bibliography{thesis}

\end{document}